\documentclass[aps,11pt,amsmath,amssymb,nofootinbib,a4paper,showpacs]{revtex4}%twocolumn
\usepackage[breaklinks]{hyperref}
\usepackage[latin1]{inputenc}
\usepackage{graphicx}
\usepackage{color}
\usepackage{amsfonts,amsthm}
\usepackage{bm,bbm}
\usepackage[OT2,OT1]{fontenc}
\newcommand\cyr{%
\renewcommand\rmdefault{wncyr}%
\renewcommand\sfdefault{wncyss}%
\renewcommand\encodingdefault{OT2}%
\normalfont
\selectfont}
\DeclareTextFontCommand{\textcyr}{\cyr}

\begin{document}
\title{Levels of spacetime emergence in quantum gravity}
\author{{\bf Daniele Oriti}}
\affiliation{Max Planck Institute for Gravitational Physics (Albert Einstein Institute) \\ Am Muehlenberg 1, D-14476 Potsdam-Golm, Germany, EU}
%\affiliation{II Institute for Theoretical Physics, University of Hamburg,  Luruper Chaussee 149, 22761 Hamburg, Germany, EU}
\email{daniele.oriti@aei.mpg.de}
%\date{\small July 30, 2014}
\begin{abstract}
We  explore the issue of spacetime emergence in quantum gravity, by articulating several levels at which this can be intended.  
These levels correspond to the reconstruction moves that are needed to recover the classical and continuum notion of space and time, which are progressively lost in a progressively deeper sense in the  more fundamental quantum gravity description. They can also understood as successive steps in a process of widening of the perspective, revealing new details and new questions at each step. 
Each level carries indeed new technical issues and opportunities, and raises new conceptual issues. This deepens the scope of the debate on the nature of spacetime, both philosophically and physically. 

\end{abstract}
%\pacs{04.60.-m, 98.80.Qc}
\maketitle
\section*{Introduction}
The problem of quantum gravity is terribly multi-faceted and can be characterized in very different ways. It is the problem of obtaining a quantum theory of geometry and spacetime, a complete quantum description of gravitational phenomena. This is common understanding: a possibly modified version of quantum mechanics should provide the mathematical framework of the theory and the object of the theory should be the gravitational interaction and the geometry of spacetime. The latter have been inextricably linked by General Relativity and nobody expects this link to be eliminated in a more fundamental quantum gravity theory. Beyond this common understanding one finds a variety of perspectives, which is moreover rapidly changing over time. This variety of perspectives, in turn, corresponds to a variety of approaches \cite{QGapproaches}. One could identify two main schemes, each comprising several specific  formalisms: one corresponding to the idea of quantum gravity resulting from quantizing a classical theory of geometry and gravity (e.g. General Relativity) \cite{QuantumGR,QuantumGR2,QuantumGR3}, the other in which spacetime, geometry and gravity are in some sense \lq emergent\rq from something else \cite{EmergentQG,EmergentQG2,oriti-emergence}. In fact, not only the distinction is very coarse grained, but it is ambiguous since the issue of the \lq emergence\rq of features of spacetime and geometry appears also in the first scheme. The emergent paradigm is the most recent and it is acquiring traction in recent years. Especially from the perspective of this second scheme, the problem of quantum gravity can be stated as: to identify the fundamental (quantum) degrees of freedom of spacetime, the \lq\lq atoms\rq\rq of space (or spacetime); to define a consistent quantum dynamics for them; to show that a continuum and classical spacetime (with a geometric and matter fields) emerges from it, in some approximation; to show that General Relativity is a good effective description of the dynamics of this emergent spacetime.

Quantum gravity in general, and the emergent paradigm in particular, face a large number of conceptual issues and raise an even larger number of philosophical questions \cite{conceptual-kiefer, QG-Phil,QG-Phil2}. This is inevitable, given the fundamental nature of the problem, shaking the very foundations of our thinking about the natural world, i.e. space and time. The (necessary and useful \cite{proliferation}) existence of a number of different approaches tackling the problem from different conceptual perspectives makes the situation more complex still. Plus, every solution is tentative, every approach is incomplete, even when solid or promising. We are truly at the chaotic frontier of knowledge. The situation for philosophical reflections is excellent. It is also very different, however, from most  philosophy of physics, since we are not dealing with the conceptual issues arising within established (mathematically and observationally) physical theories. The only way to deal with this peculiar situation is to exercise extra caution in adopting the points of view coming from specific approaches to quantum gravity as if they were more established than they are, and to refrain from resting too much on specific results as if they were necessary part of any future theory. The same situation calls for more work to map this complex landscape (see also \cite{mapQG}, especially at the conceptual level. This is what we hope to achieve with our contribution: a tentative map of the meanings in which space and time can be understood as \lq emergent\rq in quantum gravity, and of the conceptual issues associated to this emergence, and thus a greater conceptual clarity about these issues.    

The notion of emergence is itself subtle to define, even in ordinary physical theories \cite{emergence,emergence-physics-phil,emergence-physics-phil2}. We will base our analysis on a very general characterization of it, provided by Butterfield and collaborators \cite{emergence-physics-phil3, emergence-physics-phil4,emergence-physics-phil5}.  Emergence is understood to be the appearance, in a certain description of a physical system, of properties that are {\it novel} with respect to a different (more \lq fundamental\rq) description of the same system, and {\it robust} thus stable enough to represent a characterization of the new description and to form part of new predictions stemming from it. Emergence, in this understanding, usually requires the use of some {\it limiting procedure} and of a number of (possibly drastic) {\it approximations}, to allow the novel properties to become visible in the new description. 

This notion of emergence is compatible with the situation in quantum gravity, and it has been indeed already applied in this context \cite{emergence-physics-phil3, oriti-emergence, emergenceST-phys3}. Our analysis will be based on this and on a growing literature about the emergence of space and time in quantum gravity, concerning both physical and epistemological issues, among which: how to characterize this emergence and which physical consequences it may (or may not) have \cite{emergenceST-phys, emergenceST-phys2, emergenceST-phys3, emergenceST-phys4, emergenceST-phys5, emergenceST-phys6, emergenceST-phys7}, what are the ontological implications of emergent spacetime scenarios \cite{emergenceST-meta, emergenceST-meta2}, and more. Like the rest of the philosophy of quantum gravity, reflections on these issues could impact considerably, we believe, on philosophy of physics more generally, and for metaphysics and epistemology, since they challenge important aspects of these domains as well.   

The scope and content of this contribution, however, are much more limited. We will illustrate four levels of emergence for space, time and geometry (thus, the gravitational field) in quantum gravity formalisms. We discuss four ways in which space, time and geometry may be said to disappear in quantum gravity and, consequently, have to emerge to recover the description provided by General Relativity, within a more fundamental quantum gravity formalism. These four levels have to be understood as successive steps in a process of widening of the perspective, revealing new details and new conceptual issues and new questions at each step. They also represent a deepening of our understanding of the issue of the emergence of space and time in quantum gravity. They should not be misunderstood as successive, sequential ontological or inter-theoretical steps. They are not characterized each by different entities and they are not described each by a different theoretical framework. On the contrary, some of them can share the same fundamental degrees of freedom and all can be part of the same theoretical framework or quantum gravity formalism.

\section{Level 0 - Classical and quantum (modified) General Relativity}
The zeroth level of spacetime emergence is the one corresponding to the traditional idea of quantum gravity as \lq quantised GR\rq (or variations thereof). 

In the classical theory, we have a covariant set of equations for the spacetime metric (identified with the gravitational field) and matter fields living on the same differentiable manifold, following from the gravitational action of choice. These equations encode the dynamics of spacetime. 

The latter can be identified with the metric field itself or with the spatiotemporal quantities (temporal intervals, spatial distances, etc) computed out of it. Since material objects are usually required to give physical meaning to such quantities, one can instead identify spacetime with specific combinations of matter and metric fields. One could call spacetime also the differentiable manifold itself (after all this is what gives the first intuitive notion of \lq spacetime point\rq), but this is of dubious physical significance, since the dependence of physical quantities on individual points in the differentiable manifold is removed by the request for diffeomorphism invariance \cite{Carlo-diffeos}, the gauge symmetry of GR. 

Diffeomorphism invariance is indeed a key mathematical ingredient at the root of many of the conceptual difficulties about the nature of space and time in classical GR, and which have to do with the variety of possible identifications hinted at above \cite{Norton, Pooley}. 

A more physical way of characterizing these difficulties is to say that they arise from the fact that every ingredient of the theory entering the definition of \lq spacetime\rq, matter and metric fields, is {\it dynamical} and that the dynamics itself, and its generic solutions, do not select any preferred time or space direction. On the contrary, the theory admits an infinity of equally valid local notions of time and space (that could associated to specific coordinate frames, but without attributing to the latter any special physical significance). 

In this sense, one can already speak of a disappearance of space and time in classical relativistic gravitational theory \cite{carlo-time}. It is bypassed, in some sense, by the use of special solutions of the dynamics, which possess global spacetime symmetries and thus select special spacetime directions. In fact, much of gravitational physics rests on the use of such solutions.

At the conceptual level, this is already a big challenge to our customary conceptions of space and time, and raise many subtle issues, which form the subject of a vast literature in the philosophy of spacetime \cite{phil-ST, phil-ST2, phil-ST3}.    

Notice that we are not distinguishing, here, between Lagrangian or Hamiltonian formulations of the theory, even though they are not strictly speaking equivalent (diffeomorphism symmetry is implemented in subtly different ways in the two settings, and a canonical Hamiltonian formulation requires global hyperbolicity, thus it is a priori less general than the covariant, Lagrangian one). We are also not distinguishing between space and time, even though the absence of a preferred notion of time is especially troublesome for our usual understanding of physical dynamics and of physics more generally. These special difficulties are the \lq problem of time\rq in classical GR. These distinctions are not crucial to the main points we want to make in this contribution.

At the quantum level, assuming a standard formulation of quantum mechanics,  the situation is much the same, just a little worse. The kinematics has states forming a Hilbert space which encode the geometry (intrinsic and extrinsic) of spatial submanifolds (possibly forming the boundaries of the differentiable manifold) and the values and momenta of matter fields on them, and the possible histories of the same states, thus the spacetime metric and the matter fields for the whole manifold. The dynamics is encoded in some operator equation, taking necessarily the form of a constraint equation for the same data, or in a sum-over-histories, path integral formulation of the \lq transition amplitudes\rq or \lq 2-point functions\rq between the same quantum states (e.g. those interpreted as defining a physical scalar product for them), depending on the chosen classical action. Again, and for the same reasons as in the classical case, no preferred space or time direction is present in the theory, coordinate frames are unphysical and generic physical configurations of the quantum spacetime will also not select any. The situation is worse than in the classical theory because even quantum states that solve the dynamics and possess special symmetries will usually not select exact metric or matter configurations, but mostly because a preferred time direction is essential to the standard formulation and interpretation of quantum theory itself, for any physical system we know of. Thus, in the quantum gravity case we are at loss. However, this seems to us more an important problem in the foundations of quantum mechanics (can we build a consistent interpretation of quantum mechanics that does not rely, even implicitly, on a notion of time?), that any quantum  gravity theory will force us to tackle, rather than new problems with the nature of space and time themselves, which remain essentially those of the classical theory.  

At both classical and quantum levels, a solution to the problem of time and space can be found in the relational strategy \cite{relational, relational2}. It takes on board the main lesson of GR, and it rephrases it in a way that immediately suggests a tentative solution: there is no time and no space, but only physical (imperfect) clocks and rods. The strategy amounts to identifying internal degrees of freedom of the complete system composed of metric and matter fields that can be used as {\it approximate} rods and clocks to parametrize the spatial relations and temporal evolution of the remaining degrees of freedom. To us, this is an adequate solution to the issue of defining space and time in a (quantum) relativistic context, and a very physical one (but for a sample of the remaining issues, see \cite{relational2, relational3, relational4}). It forces us, however, to accept the fact that physical clocks and physical rods will never be perfect, i.e. matching the idealized (but unphysical) notion of time  and space provided by coordinate systems. This is simply the other inevitable side of their being physical systems: quantum and interacting with the ones they parametrize.

There is thus a sense in which space and time disappear in classical General Relativity and, in a more drastic sense, in the quantum General Relativity. There is thus also a sense in which space and time have to {\it emerge} also in this context. In the classical case, this amounts to the dynamical selection of symmetric spacetimes or to the approximation leading to physical rods and clocks behaving as perfect ones. In the quantum case it is part of the standard problem of the classical approximation of a quantum theory, since the above symmetric spacetimes or geometries, and the close-to-ideal clocks and rods have to emerge from ultimately quantum entities. General covariance, once more, leads to several additional complications to this notoriously already difficult problem, but, we maintain, does not change its nature. Most of the above challenges are well explored in the quantum gravity literature. But this is only Level 0 of spacetime emergence.

\section{Level 1 - New degrees of freedom - Geometry and spacetime as emergent entities}
An altogether different sense in which space and time disappear in quantum gravity, and thus have to emerge in some approximation, is central in quantum gravity approaches that do not deal simply with quantized gravitational and matter fields. A new level in reached when quantum gravity formalisms are based on new types of quantum degrees of freedom which are not geometric in a straightforward way, but of a different nature, usually combinatorial and algebraic. In particular, this often implies a fundamental discreteness of the same quantum entities. The spin network states of loop quantum gravity \cite{LQG,LQG2,SF}, with their dual functional dependence on group elements or group representations associated to graphs, and their histories labeled by the same algebraic data and associated to cellular complexes, fit this characterization\footnote{This is true even though, historically, they have been \lq discovered\rq within a rather conservative strategy of quantizing the gravitational field once it has been recast in the language of gauge theories.}. The simplicial (piecewise-flat, thus singular) geometries of lattice quantum gravity approaches like quantum Regge calculus \cite{qRC} and (causal) dynamical triangulations \cite{CDT} can also be understood in this perspective. The quanta of group field theories \cite{GFT,GFT2}, which can be described both as generalised spin networks and as simplicial building blocks of piecewise-flat geometries, and whose quantum dynamics merges the idea of spin foam models and that of lattice quantum gravity, are another example. Causal sets \cite{CS} are another purely discrete replacement for continuum fields. String theory offers a number of results all pointing to the replacement of the notion of continuum geometric fields as fundamental entities \cite{ST-emergent} and to a much more general type of geometry being reconstructed from the dynamics of strings \cite{ST-geometry}. Other examples could be cited. The main point should be clear: in quantum gravity, the fundamental diegrees of freedom are not continuum fields and spacetime dissolves into pre-geometric, non-spatiotemporal entities, from which space, time and geometry have to emerge in some approximation.

With the appearance of new fundamental (quantum) entities replacing continuum fields, call them \lq atoms of space\rq, an altogether new dimension opens up for quantum gravity research. Beside identifying the properties and dynamics of such fundamental entities, the crucial task becomes to understand by which physical mechanisms and under which approximations they become amenable to a description in terms of continuum spacetime and geometry (and matter fields). This is the problem of the continuum limit in discrete quantum gravity approaches. It must be carefully  distinguished, conceptually and mathematically, from the classical limit mentioned in the previous section. It rests on coarse graining and renormalization schemes, the identification of appropriate collective observables, and in particular on the identification of the usual continuum fields (metric, matter fields) as examples of collective quantities built out of the more fundamental atoms of space. 

It is along this new dimension that we expect space and time to emerge in a stronger sense from entities that are not spatiotemporal. The new physics leading to the emergence of spacetime, more precisely, can be expected to be the one captured by increasing numbers of interacting, quantum fundamental  entities, with space, time and geometry arising from the collective behaviour of the same. Collective behaviour is indeed the prototypical producer of emergent physics, and the conceptual setup would here sees spacetime in analogy with condensed matter and quantum mnay-body systems \cite{ST-manybody}. With increasing numbers of fundamental entities, there come new emergent properties, new approximations and effective dynamics, and with them, new concepts.

The above expansion of the scope and content of quantum gravity research has been discussed in some detail in \cite{Bronstein}, including also a brief survey of recent developments. These include spin foam lattice renormalization \cite{SFrenorm,SFrenorm2,SFrenorm3,SFrenorm4,SFrenorm5}, continuum limits in random tensor models \cite{TM,TM2} and dynamical triangulations \cite{CDT}, group field theory renormalization \cite{GFTrenorm,GFTrenorm2,GFTrenorm3,GFTrenorm4,GFTrenorm5,GFTrenorm6}, the extraction of effective cosmological dynamics as group field theory hydrodynamics \cite{GFTcosmo,GFTcosmo2,GFTcosmo3,GFTcosmo4,GFTcosmo5,GFTcosmo6}.

Here, we want to stress the conceptual shift that such expansion brings, concerning the nature of space and time. When looked at from the point of view of the (candidate) pre-geometric atoms of space (level 1), it is clear that space, time, geometry, matter have dissolved in a deeper, more radical sense than from the perspective of quantised GR (level 0). There, one had to get rid of the idea of any preferred notion of space and time, and a multiplicity of potential physical notions (be them defined by special configurations of spacetime geometry or by relationally defined frames). All such notions were made possible and had to be constructed by means of continuum fields. They are now missing. Thus, even the {\it possibility of space and time}, in the sense of level 0, has yet to emerge. It has to be obtained by moving along the new dimension of increasing numbers of fundamental building blocks and by exploring their collective properties.

This raises a number of questions concerning the nature of the atoms of space themselves. In particular, to what extent do they carry spatiotemporal properties at all? The need to reconstruct space and time from them, at least in some approximation and with respect to special aspects of their collective dynamics, implies that they carry at least \lq\lq seeds\rq\rq of space and time with them. By this we mean that some of their properties should be {\it translatable} into spatiotemporal notions at least in those approximations, even though they are not fully spatiotemporal in general. In other words, if spacetime has to be reconstructed at all, the more fundamental theory should allow for a dictionary, mapping its basic entities and some of their properties into continuum fields including those defining spatiotemporal notions, in some sector of the same theory and in an approximate manner. The map will certainly not be one to one, nor exact, but it should exist if the candidate fundamental theory should have any physical relevance at all. The existence of such dictionary, i.e. being part of the domain of this translation map, implies a \lq proto-spatiotemporal\rq characterization of some properties of the fundamental atoms of space (and justifies their name). Nothing more than that should be assumed, however. 

A more precise characterization requires probably to consider specific examples of candidate atoms of space and of quantum gravity formalisms. In particular, it is possible that some properties attributed to such atoms of space, among those that are crucial in reconstructing the standard notions of space and time, can be understood as offering a more primitive notion of space and time (e.g. based on adjacency, ordering, etc), farther away from usual physics, but arguably more fundamental. A more primitive spatiotemporal reality would then replace, despite its radical departure from any traditional understanding (and use) of space and time, the one that we are accustomed to. This may end up being simply a matter of nomenclature. If the new properties are truly radically different (in mathematical and physical understanding) from the space and time of continuum relativistic physics, to call them \lq spatiotemporal\rq may not offer more than a psychological relief. 

An important issue is the ontological nature of the new fundamental entities underlying spacetime and, conversely, of space and time themselves, once we deprive them of their fundamental status and understand them as emergent. In fact, modern ontology \cite{ontology} is based explicitly or implicitly on spatiotemporal notions, to the point that \lq to be real\rq is often thought equivalent to \lq to exist in space and time\rq , i.e. to have a well-defined location and stable duration. This already raises ontological issues concerning the fields (in particular the metric field) that are used to define location and duration in relativistic physics. But the same ontological issues are brought to a whole new stage when referred to the putative atoms of space, underlying the same continuum fields and replacing them at the fundamental level. Conversely, unless one adopts the radical opposite of the usual position (to be real is to exist in space and time) and thus deprive space and time of {\it any reality at all}, due to their loss of fundamental status, one is forced to revise the very notion of reality in presence of emergent behaviour. One has to accept a multi-level ontology of some sort, in which both fundamental and emergent properties and entities are real in an appropriate sense. In other words, an emergent spacetime scenario forces a radical revision of metaphysics in parallel with the revolution in physics that it represents, concerning what is meant by {\it real} (which has to be independent to some extent from spatiotemporal properties) and what this attribute is assigned to (which probably has to be done in a more liberal and less exclusive way). For recent work on these issues, see \cite{emergence-physics-phil3, emergence-physics-phil4,emergence-physics-phil5,emergenceST-meta,emergenceST-meta2}.

Another set of issues raised by emergent spacetime scenarios is of a more epistemological nature. It concerns the physical salience of the candidate atoms of space and of the the theories describing them, and their empirical coherence \cite{emergenceST-phys3,emergenceST-phys4,oriti-emergence}. The worry is that, because we live in spacetime and the notions of space (e.g. locality) and time (e.g. duration) are at the very root of our empirical access to reality, any theory formulated without them is either empirically empty or empirically incoherent. We maintain \cite{oriti-emergence} that the necessary requirement of reproducing some (possibly modified) form of relativistic spacetime physics settles the worry of empirical emptiness of emergent spacetime scenarios, at least as a matter of principle. We also maintain that the empirical coherence of the same scenarios will have to be ensured by the details of such spacetime reconstruction, and of course tested in each specific formalism, but again that there is no obstruction in principle. The conceptual difficulties of course remain, and have to be consistently and seriously tackled in any quantum gravity formalism. We refer to recent literature for more details \cite{emergenceST-phys4,emergenceST-phys7}.

To summarize, the existence of new types of physical degrees of freedom, of a non-spatiotemporal type (in particular, different from continuum quantum fields), suggested by several quantum gravity approaches, points to an emergent nature of space, time and geometry (and matter). It enlarges greatly the scope of quantum gravity research by requiring a focus on such emergence (which includes the continuum limit of the fundamentally discrete quantum gravity structures) and by raising a large number of conceptual issues. These include both ontological questions about the nature of spacetime and of its more fundamental \lq constituents\rq, and epistemological questions about their empirical significance and accessibility. 

Notice that, while the technical issues related to the emergence of spacetime in such quantum gravity approaches are not much affected by the interpretation of the \lq atoms of space\rq , the conceptual issues listed above certainly are.  Even if we regard them as mere technical tools encoding some sort of regularization or representing simply an intermediate step towards the true definition of the theory in terms of quantized continuum fields, the problem of the continuum limit remains the key one to tackle, via coarse graining and renormalization, as it remains necessary to devise observables that encode continuum physics in terms of the discrete building blocks one uses at first. In this case, however, no new conceptual issue arises with respect to level 0, since no meaning is assigned to any part of the theory before the same continuum limit is taken and the formulation of the theory in terms of continuum quantum fields is achieved. Not so if we give a realistic interpretation to the atoms of space suggested by the formalism and thus we investigate their physics and metaphysics even before spacetime has emerged.

\section{Level 2 - Non-geometric phases - The atoms of space(time) are really not spatio-temporal}
If we take a realistic stance towards the non-spatiotemporal atoms of space, we should be ready for even more conceptual challenges. 

Moving along the direction of increasing numbers of them, thus exploring their collective behaviour and continuum limit, we should expect to find that the continuum limit of such system is not unique. This is what generally (there are of course exceptions, which we treat as such) happens for any system of many interacting quantum degrees of freedom. The quantum dynamics of such interacting systems leads normally to different macroscopic phases, separated by phase transitions. Each macroscopic phase is characterized by different emergent properties, different macroscopic observables and a different effective dynamics. In some sense, the underlying microscopic quantum system is \lq replaced\rq by a very different kind of emergent, macroscopic system in each phase. A different macroscopic phase is, in many ways, a \lq different world\rq \cite{Strocchi}.

For our non-spatiotemporal, quantum gravity system of atoms of space, the key issue becomes then to identify such macroscopic phases and, among them, the one (or more) in which an effective description in terms of space, time and geometry is possible, and it is governed by an effective general relativistic dynamics, at least in some approximation. In other words, the emergence of spacetime that we envisaged in the previous section should be expected to take place only in one (or some) of the possible macroscopic phases, in which the fundamental non-spatiotemporal atoms of space organize themselves. It is the task of quantum gravity formalisms that suggest fundamental non-geometric atoms of space to show that there exists such geometric, spatiotemporal phase, in a continuum limit, in some approximation.

Quantum gravity approaches have embraced this task and have obtained considerable progress in recent years. New phases (alternative to the usually non-geometric ones in which the formalisms are first defined) have been studied in the loop quantum gravity context \cite{BF-phase,BF-phase2,BF-phase3,tim,timhanno} and in group field theory \cite{GFTphases}, where condensate states have also been put in correspondence with effective cosmological dynamics \cite{GFTcosmo,GFTcosmo2,GFTcosmo3,GFTcosmo4}. Indications of phase transitions have been obtained in spin foam models \cite{SFrenorm,SFrenorm2,SFrenorm3,SFrenorm4} and again in the group field theory context \cite{GFTrenorm,GFTrenorm2,GFTrenorm3,GFTrenorm4,GFTrenorm5}. Extensive studies of the phase diagram of simplicial quantum geometries, and supporting evidence for an extended De Sitter-like geometric phase are at the core of causal dynamical triangulations \cite{CDT-phases}. Similar work has started recently in the causal set programme \cite{Lisa-CS-phases}. More examples could be mentioned. 

One important note concerns the coupling constants, or other parameters, which characterize the quantum gravity phase diagram. In simplicial quantum gravity approaches, they are usually identified directly with the same coupling constants of continuum gravitational theories (Newton's constant, cosmological constant, etc), since the quantum dynamics is defined as a discretization of this. In spin foam models, the situation is similar, but new parameters may enter, motivated by the specific model building guidelines or by the renormalization schemes. In group field theories, as well as in tensor models, the matching with gravitational dynamics is searched for only in a continuum approximation (even if it can be in principle performed also at the discrete level, as in simplicial quantum gravity). This matching is anyway a necessity in any quantum gravity formalism. 

There is a clear sense in which, in the presence of a non-trivial macroscopic phase diagram, thus of different phases only some of which (hopefully) of a spatiotemporal and geometric nature, spacetime and geometry can be said to be {\it emergent} and not fundamental in a deeper and more radical sense that exposed at Level 1. 

Of the new issues raised by the existence of non-spatiotemporal atoms of space, mentioned in the previous section, the epistemological ones are not much affected. The ontological ones are. 

The reason is that such atoms of space are now deprived even more of any spatiotemporal attribute, even though they remain, mathematically, the very same entities identified at Level 1. In fact, whatever properties of such entities end up producing spatiotemporal observables or dynamics (e.g. some \lq volume/extension attributes\rq), after coarse-graining or some other approximation, or after being treated in a collective manner, they do so {\it only} in some specific phase of the system (e.g. only for specific values of the coupling constants or macroscopic parameters characterizing it). They may be \lq seeds\rq of an emergent spacetime, in some sense, but one is precluded the possibility to consider them \lq spatiotemporal properties\rq in disguise. Yet in other words, something more radical is at play than a simple \lq approximation\rq . For example, continuum spacetime and geometry are not just an approximate construction from a discrete and quantum spacetime and geometry, differing only in some aspects but sharing the same nature. The very same entities, even when looked at in the same macroscopic approximation or treated by analogous coarse-graining techniques, may not produce a continuum spacetime or geometry at all. Their ontology has to be understood as being of a truly different kind. In parallel with it, the ontological status of continuum spacetime and geometry has also to be understood differently, since it turns out to be emergent in an even more radical sense.

Of course, new issues arise also at the epistemological level, at least in the sense that many of the same questions raised at Level 1 have to be further refined in the presence of new quantum gravity phases for our universe. Any further analysis of such epistemological refinements, as well as of the new ontological issues raised at this new level, will have to be carried out in the context of specific quantum gravity formalism. 
In any case, the very existence of such new issues is the reason to emphasize the existence of such new level of spacetime disappearance (and emergence). 

\section{Level 3 - Geometrogenesis - The emergence of spacetime via a phase transition as a physical process}
The process of deepening and broadening our perspective on spacetime disappearance and emergence, stimulated by the hypothesis of new non-spatiotemporal entities underlying the universe, proceeds even further once the possibility of new macroscopic (continuum) phases is granted. If a realistic interpretation of the fundamental atoms of space is valid, and they can organize themselves in different collective phases, there is no obvious reason why the phase transition separating non-geometric from geometric phases should not be regarded as physical as well.  

This phase transition, dubbed \lq geometrogenesis\rq , and the mechanisms producing it, has been studied in a number of quantum gravity formalisms. From a more physical perspective, it has been first discussed in \cite{Qgraphity} in a graph-based approach to quantum gravity, and immediately afterwards in the group field theory formalism \cite{GFTfluid}. More recently, it has been discussed in relation to the phase diagram of causal dynamical triangulations as well \cite{jacub-geometrogenesis}. 
Its conceptual aspects, on the other hand, have received little attention until now (see \cite{emergenceST-phys7} for recent work).

Investigations of such conceptual aspects, however, will have to rely on a better understanding of the physical nature of the geometrogenesis phase transition as a physical process. But what type of physics does it capture? 

One natural hypothesis is that it should be given a cosmological interpretation, as the process that underlies (or replaces) the big bang, as the origin of the physical universe as described by General relativity and quantum field theory. This is the suggestion made in the mentioned studies of geometrogenesis, and it has also been explored from a tentative phenomenological perspective in a cosmological context in \cite{geometrogenesis-cosmo}. It is also the underlying hypothesis of GFT condensate cosmology \cite{GFTcosmo,GFTcosmo2,GFTcosmo3,GFTcosmo4,GFTcosmo5}, where geometrogenesis is technically implemented as a {\it condensation} of the microscopic atoms of space, with the emergent universe described in analogy with a quantum fluid\footnote{In the GFT cosmology context, the idea of geometrogenesis as replacing the big bang competes with the alternative idea of a bouncing scenario, as discovered in the simplest hydrodynamic description of the system.}. It resonates as well with the so-called emergent universe scenario for primordial cosmology, an alternative to cosmic inflation, first proposed in \cite{emergent-universe} and also realised in the context of string gas cosmology \cite{string-gas-cosmology}\footnote{Here, however, the emergence process does not involve the temporal aspects of the universe, since a time direction remains well defined during the whole cosmic evolution, even across the phase transition.}. 

While this cosmological interpretation is suggestive and, indeed, natural, it is also tricky and at risk of misunderstanding. The main difficulty is the immediate temptation to interpret a cosmological phase transition not only as physical but also as a {\it temporal} process. This is also a problem with the very language we use to characterize physical {\it processes}. A phase transition is pictured as the outcome of \lq evolution\rq in the phase diagram of the theory, or of a \lq flow\rq of its coupling constants; we say we \lq move\rq towards the cosmological, geometric phase from the non-geometric, non-spatiotemporal phase, or viceversa. However, we are dealing with a system which is already described at Level 2: there is no continuum space, no continuum time, no geometry in the usual sense; and it is also not characterized by features which are just \lq\lq one approximation away\rq\rq from time and space.

So, first, we need to have a background-independent and non-spatiotemporal notion of \lq evolution\rq in the space of quantum gravity coupling constants, i.e. in the \lq theory space\rq characterizing the quantum gravity formalism at hand. Notice that such evolution will relate different continuum theories, in particular different macroscopic effective dynamics, for the same fundamental quantum entities. This notion of evolution in theory space is what specific renormalization group schemes in various quantum gravity formalism will provide.

Next, we can ask whether any notion of \lq proto-time\rq and \lq proto-temporal\rq evolution can be associated with such flow in a quantum gravity theory space, and how it relates to any of the notions of time that may emerge in the geometric, spatiotemporal phase of the universe (thus, \lq after\rq the geometrogenesis phase transition). 

There are two orders of difficulties here. One is the mentioned absence of any notion of time at Level 2, which adds conceptual difficulties to the absence of any notion of time of Level 1, and to the \lq problem of time\rq in (quantum) GR, i.e. Level 0. The other is that, strictly speaking, even the standard RG flow of coupling constants in ordinary statistical (field) theory is \lq timeless\rq and not interpreted as standard evolution, since it may well refer to systems {\it at equilibrium}\footnote{Of course, when one is dealing with systems out of equilibrium, and thus time-dependent, this additional difficulty is absent.}. The reason why we have no particular conceptual issue in understanding the flow in theory space and the approach to phase transitions in temporal terms, despite the fact that they refer to a change in the time-independent coupling constants of the system, is that we can easily imagine an external observer (the experimental physicist in the lab) tuning such coupling constants towards their critical values, and thus pushing the system towards the relevant phase transition. Needless to say, no such external observer is available in quantum gravity.

Any notion of time or, better, \lq proto-time\rq that could be associated to such flow across the quantum gravity phase diagram would in any case deserve such name only in the sense that, once used to parametrize the flow across a non-geometric phase towards a geometrogenesis phase transition, it ends up matching some spatiotemporal observable that can be used as a time variable within the geometric phase. Viceversa, it would correspond to what is left of some geometric variable used to define a notion of time in such phase, and used as well as a notion of RG scale for the quanutm gravity system, once the same system flows across a geometrogenesis phase transition into a non-spatiotemporal phase.

We leave a detailed and concrete analysis of this problem, and of the many conceptual issues associated to it, to future work. Here, we only suggest a possible strategy, which can be understood as \lq pushing the relational framework (used to obtain a notion of time at Level 0) two levels forward\rq. The idea would be to take an internal (dynamical) degree of freedom, used as a relational clock in the geometric GR-governed phase at Level 0, to parametrize (as the relevant notion of \lq scale\rq) the RG flow of the underlying non-spatiotemporal quantum gravity system. This would allow to give a proto-temporal evolution interpretation to the same RG flow, even in the non-geometric phases, and thus to the geometrogenesis phase transition. One example could be the emergent (free, massless) scalar field used in group field theory cosmology \cite{GFTcosmo4,GFTcosmo5,GFTcosmo6} as well as in (loop) quantum cosmology \cite{LQC};. Another could be any observable playing the role of \lq volume\rq or scale factor of the universe, in the geometric phase; one could imagine then the RG flow to drive the system and its coupling constants from large volumes towards very small volumes and there hitting a geometrogenesis phase transitions (thus replacing the big bang singularity), ushering the universe into the non-spatiotemporal phase (where the interpretation of the same variable as \lq volume\rq or relational time would cease, contextually, to make sense).

A scenario of the type sketched above, we believe, will require serious reflections not only on the nature of space and time, but also on the renormalization group when applied to quantum gravity and on a covariant, spacetime-free understanding of statistical mechanics. Even more clearly, it may have profound implications on the philosophy of cosmology, that will be subject to a broadening of scope and perspective in parallel with the one we are suggesting for the philosophy of quantum gravity.

In the end, these many new conceptual issues that arise in this scenario are the reason to associate to it a new level of spacetime disappearance and emergence.

\section{An analogy: Bose-Einstein condensates}
Before concluding, we would like to offer a physical analogy of the situation outlined for quantum gravity, and of the various levels of \lq emergence\rq we illustrated in this contribution. We hope this will clarify further the conceptual framework we have in mind. For more details on this example, see \cite{volovik-QG-hydro}

Consider the hydrodynamic description of a fluid, with the main dynamical variables being the fluid density and velocity, and interesting observables being the total momentum and energy, vorticity, circulation of vortex excitations, viscosity, etc, which are functions of them. On top of the global configurations of the fluid, one has propagating excitations over them corresponding to sound waves with their own characteristic dispersion relation.
Notice that one can also consider extended versions of standard fluid hydrodynamics, including new terms functions of the same density and velocity fields (e.g. gradient terms); this is the case, for example, of superfluid hydrodynamics. In our analogy, standard hydrodynamics would be the counterpart of GR, with spatiotemporal, geometric observables (volumes, areas, distances, time intervals, curvature, etc) corresponding to various hydrodynamic observables, functions of the basic fields (in GR, the metric or metric and connection etc, plus matter fields, depending on the specific formulation of classical gravity one uses). 

From this classical theory, one can move to the quantum regime. One could start by simply quantizing the classical theory. The resulting quantum theory is perturbatively non-renormalizable as a quantization of sound waves (in fact, with the same order of divergences as perturbative GR in terms of gravitons) but it could make sense non-perturbatively. In the presence of gauge symmetries, these will carry over to the quantum level and require observables to satisfy them (the analogue of the problem of time in quantum gravity). Emergent notions of time and space associated to observables constructed via the relational strategy correspond, in the analogy, to specific hydrodynamic observables, also at the quantum level but thus subject to extra complications due to their quantum nature. The same story goes for superfluid hydrodynamics. This is our Level 0.

We know that there is a deeper level of description (and of physics) for fluids, which becomes even more relevant for superfluids: the underlying quantum theory of atoms. This could be described by a  quantum field theory for bosonic or fermionic entities, and we know that its features are not captured by the straightforward quantization of classical hydrodynamics. The microscopic quantum dynamics differs from the one corresponding to quantum hydrodynamics (even when expressed in the same variables, as in the case of superfluids described in terms of a single complex scalar field); moreover, the classical, long-wavelength approximation of quantum hydrodynamics does not reproduce the classical theory one started from, in general. For superfluids, moreover, we know that the quantum statistics of the atoms is required in a complete understanding of the key macroscopic features, i.e. superfluidity itself.  The macroscopic parameters are understood as functions of the microscopic coupling constants, e.g. the mass of the atoms. These are instances of the fact that a continuum approximation built on many fundamental degrees of freedom is conceptually and physically differnet from the simple classical approximation of the same degrees of fredom, a point emphasized in \cite{Bronstein} in the case of quantum gravity. The very basic variables of the macroscopic theory, the hydrodynamic variables, are also understood to be useful expressions of collective (and coarse-grained) properties of the fundamental atoms, and the same holds of course at the level of observables. In particular, it holds for those observables that, in the analogy, correspond to spatiotemporal and geometric quantities. Spacetime is emergent, then, in the same sense in which the hydrodynamic description is emergent from the more fundamental atomic description, based on a different kind of degrees of freedom, thus a different ontology. This is Level 1.

Broadening further the picture of the fundamental atomic system, we know that the fluid description in terms of hydrodynamics and, more specifically, of superfluid hydrodynamics, is only available in one of its macroscopic, continuum phases, the one in which it behaves indeed as a liquid. The same system can organize itself also as a gas or a solid (more exotic condensed matter systems may have an even richer phase diagram). In these other phases, the observables that correspond, in our analogy, with spatiotemporal or geometric observables will simply not be available, in general. The very hydrodynamic variables used to construct them, and analogous to the spatiotemporal continuum fields of GR, will not be available. Notice that this is an additional, more radical point than the noted difference between the two hydrodynamic descriptions of ordinary fluids and superfluids, i.e. the two different effective (and emergent) dynamics of the same atomic system in two different fluid phases. The latter may correspond, in the analogy, to two different gravitational theories or effective spatiotemporal dynamics in two different geometric phases of a fundamental non-spatiotemporal system of \lq atoms of space\rq , with physically distinguished properties. If it was already clear that atoms were not simply \lq pieces of fluids\rq or \lq fluid-like\rq in any sense at Level 1, now that we notice that they can form solids or gases or other non-fluid phases, we are truly forced to think of them as possessing a different, independent nature. This is Level 2.

There is also no doubt, in the case of ordinary fluids and superfluids, that the phase transitions separating different macroscopic phases, and in particular the fluid one(s) and the non-fluid ones, correspond to physical processes. The analogues of our additional questions concerning spacetime in the presence of physical phase transitions of our fundamental quantum gravity system would go roughly as follows. We would ask what happens to specific hydrodynamic observables across the phase transitions and whether there is any remnant of them, or simply a corresponding observable, constructed in the same way out of the atomic degrees of freedom, in the non-fluid phases. We would then try to use them to characterize the physics of the phase transition itself. We could also ask whether the same observables illuminate aspects of the RG flow across the phase diagram, i.e. the running of atomic coupling constants, and in particular if they could be used as the relevant \lq scale\rq parametrizing the same flow. In any case, when studying the meaning and physics of phase transitions of the non-fluid-like atomic system, while keeping the attention on some hydrodynamic observables, those corresponding in the analogy to spatiotemporal and geometric quantities, we would be working at Level 3.

\section*{Conclusions}
We  have explored the issue of spacetime emergence in quantum gravity, by articulating several levels at which this emergence (correspondingly, the disappearance of space and time) can be intended.  As we have shown, these levels correspond to successive steps in a process of widening of the perspective on the nature of space, time and geometry, revealing new questions at each step. 
We have highlighted some of the new technical issues, that arise at each level, and discussed in particular the new conceptual issues. The main goal was to clarify the broader perspective one can take towards the issue of the emergence of spacetime in quantum gravity, and forming the basis for a lot of recent work in the field. We believe that this broader perspective deepens the scope of the debate on the nature of spacetime, both philosophically and physically. Finally, we hope that it could be the starting point of many new research directions in the philosophy of quantum gravity and of spacetime,for what concerns both its metaphysical and epistemological aspects.  

\renewcommand{\refname}{Bibliography}

\end{document}